\documentclass[a4paper]{jpconf}

\usepackage{amsmath}
\usepackage{amssymb}
\usepackage{graphicx}
\usepackage{cite}
\usepackage{multirow}
\usepackage{longtable}
\usepackage{lscape}
\usepackage{subfigure}

\def\bea{\begin{eqnarray}}
\def\eea{\end{eqnarray}}
\def\nue{{\nu_e}}

\def\numu{{\nu_{\mu}}}
\def\anumu{{\bar\nu_{\mu}}}

\newcommand{\eg}{{e.g.}}

\newcommand{\tet}{\theta_{13}}
\newcommand{\ldm}{\Delta m_{31}^2}

\newcommand{\deltacp}{\delta_{\mathrm{CP}}}
\newcommand{\stheta}{\sin^2 2 \theta_{13}}

\newcommand{\capdef}{}
\newcommand{\mycaption}[2][\capdef]{\renewcommand{\capdef}{#2}\caption[#1]{{\footnotesize #2}}}

\begin{document}

\begin{flushright}
EURONU-WP6-11-40 \\
IFIC/11-60
\end{flushright}

\title{Optimized Neutrino Factory for small and large $\theta_{13}$}

\author{Sanjib Kumar Agarwalla}

\address{Instituto de F\'{\i}sica Corpuscular, CSIC-Universitat de Val\`encia, \\ Apartado de Correos 22085, E-46071 Valencia, Spain}

\ead{Sanjib.Agarwalla@ific.uv.es}

\begin{abstract}
Recent results from long baseline neutrino oscillation experiments point towards a non-zero value of $\tet$ at around $3\,\sigma$ confidence level.
In the coming years, further ratification of this result with high significance will have crucial impact on the planning of the future long baseline 
Neutrino Factory setup aimed to explore leptonic CP violation and the neutrino mass ordering. In this talk, we discuss the baseline and energy optimization 
of the Neutrino Factory including the latest simulation results on the magnetized iron neutrino detector (MIND) in the light of both small and large $\tet$. 
We find that in case of small $\tet$, baselines of about 2500 to 5000 km is the optimal choice for the CP violation measurement with $E_{\mu}$ 
as low as 12 GeV can be considered. However, for large $\tet$, we show that the lower threshold and the backgrounds reconstructed at 
lower energies allow in fact for muon energies as low as 5 to 8 GeV at considerably shorter baselines, such as Fermilab to Homestake.
This suggests that with the latest MIND simulation, low- and high-energy versions of the Neutrino Factory are just two different forms of the same experiment 
optimized for different regions of the parameter space.

\end{abstract}

{\it\small Contribution to NUFACT 11, XIIIth International Workshop on Neutrino Factories, Super beams and Beta beams, 1-6 August 2011, CERN and 
University of Geneva (Submitted to IOP conference series)}

\section{Introduction and Motivation}
\label{sec:intro}

The performance of future long baseline neutrino oscillation experiments to unravel leptonic CP violation (CPV) and the neutrino mass 
hierarchy (MH) is crucially dependent on the achievable event rates and hence strongly on the size of 
$\theta_{13}$~\cite{Bandyopadhyay:2007kx,Mezzetto:2010zi,Agarwalla:2010hk,Agarwalla:2011hh}.
Recent T2K~\cite{Abe:2011sj} and MINOS~\cite{Adamson:2011qu} data have provided new information on $\tet$, indicating that 0.01 $\lesssim \stheta \lesssim$ 0.1 
at $2\,\sigma$ confidence level with a best-fit value of 0.051 (0.063) for normal (inverted) mass ordering~\cite{Schwetz:2011zk}.
It is expected that in the next few years, this hint will be validated with high significance with more informations coming from the T2K, MINOS and 
upcoming reactor and NO$\nu$A experiments, opening up the possibility of observing CPV in the lepton sector if the Dirac CP phase, $\deltacp$, is not equal to 
$0^{\circ}$ or $180^{\circ}$. 

The prospects of the present generation long baseline experiments, T2K and NO$\nu$A in probing CPV and the mass ordering are very limited 
even if $\tet$ is large~\cite{Huber:2009cw}. The experiment class, which can ultimately address these unsolved issues with unprecedented 
sensitivity for both small and large $\tet$, is the Neutrino Factory~\cite{Bandyopadhyay:2007kx,Agarwalla:2010hk}.
In a Neutrino Factory, the neutrino beam is created from the decay of muons in flight in the straight sections of a storage ring. 
The International Neutrino Factory and Superbeam Scoping Study~\cite{Bandyopadhyay:2007kx,Abe:2007bi,:2008xx} has laid the
foundations for the currently ongoing Design Study for the Neutrino Factory (IDS-NF)~\cite{ids} which has come up with a 
first-version baseline setup of a high energy neutrino factory (HENF) with $E_\mu=25 \, \mathrm{GeV}$ and two baselines 
$L_1 \simeq 3 \,000 - 5\, 000 \, \mathrm{km}$ and $L_2 \simeq 7 \, 500 \, \mathrm{km}$ served by two racetrack-shaped 
storage rings~\cite{Huber:2006wb}.

The most promising avenue to explore CPV and $sgn(\ldm)$ at a Neutrino Factory is the sub-dominant $\nue \to \numu$ oscillation channel
which produces muons of the opposite charge (wrong-sign muons) to those stored in the storage ring and these  
can be detected with the help of the charge identification capability of a magnetized iron neutrino detector (MIND)~\cite{Cervera:2000kp}.
As the design of the Neutrino Factory matures, more realistic detector simulation and analysis technique have been developed
which improves the efficiency of MIND at low energies and also lowers the threshold~\cite{Cervera:2010rz,ThesisLaing}.
These new simulations provide the full response matrices for all the signal and background channels. 
In the most recent analysis~\cite{ThesisLaing}, low energy neutrino signal events down to $1\,\mathrm{GeV}$ were selected with an
efficiency plateau of $\sim$60\% for $\numu$ and $\sim$80\% for $\anumu$ events starting at $\sim$$5\,\mathrm{GeV}$, while
maintaining the background level at or below 10$^{-4}$. For the neutral current background, the impact of migration is non-negligible
and it is peaked at lower energies. This feed-down is the strongest effect of migration and thus has potential impact on the energy 
optimization, since it penalizes neutrino flux at high energies, where there is little oscillation but a large increase in fed-down background.
Now we discuss the simultaneous optimization of baseline and muon energy in a green-field Neutrino Factory scenario~\cite{Agarwalla:2010hk} 
using the new migration matrices for MIND detector~\cite{ThesisLaing}.

\section{Optimization of a green-field Neutrino Factory setup with new MIND detector}
\label{sec:LvsE}

\begin{figure}[tp] 
\centering
\includegraphics[width=13.0cm, height=10.0cm]{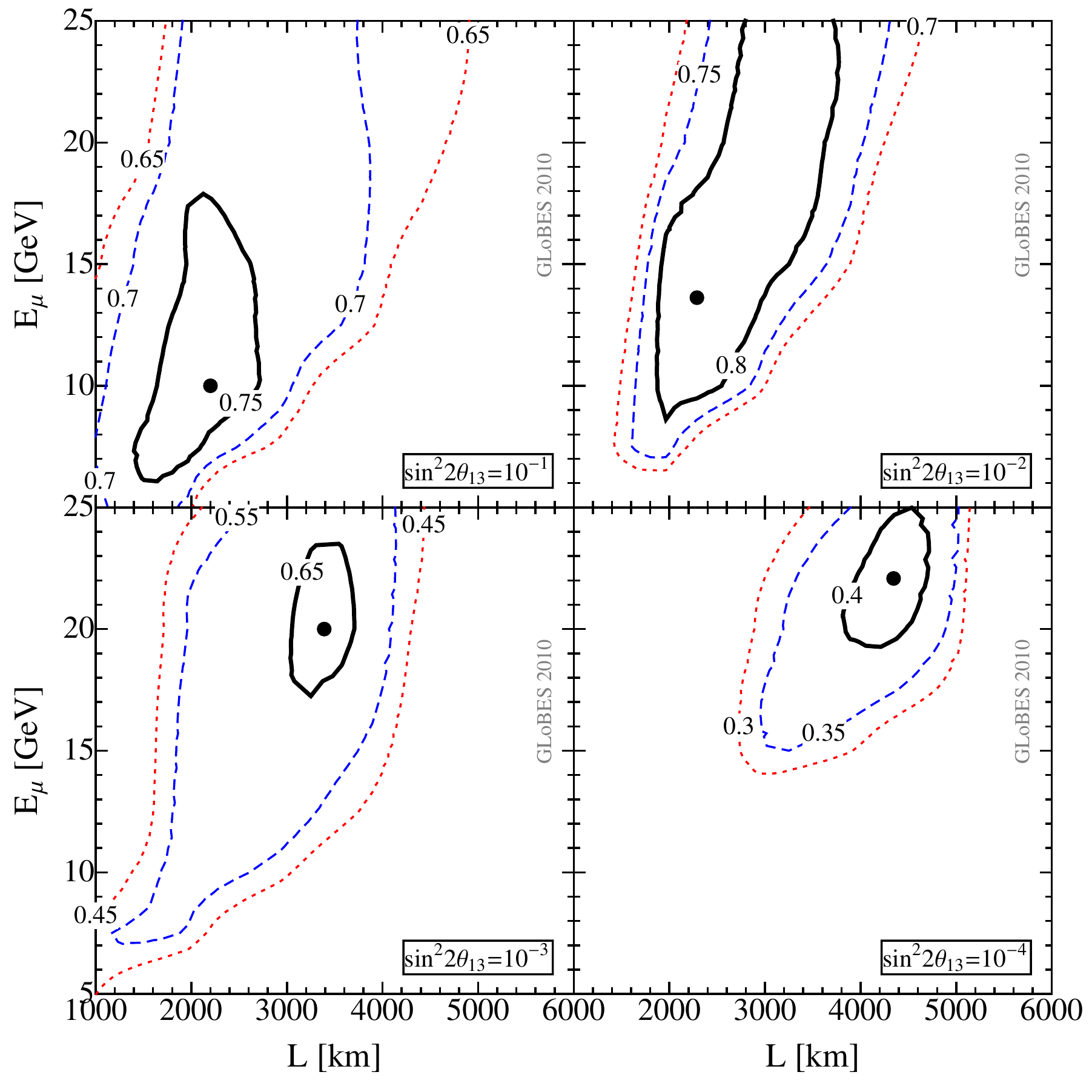} 
\mycaption{Fraction of $\deltacp$(true) for which CPV will be discovered at $3\,\sigma$ C.L. as a function of $L$ and $E_\mu$ for the single 
baseline Neutrino Factory. The different panels correspond to different true values of $\stheta$, as given there. 
Here we consider $5 \times 10^{21}$ useful muon decays in total with a 50 kt detector. The optimal performance 
is marked by a dot: (2200,10.00), (2288,13.62), (3390,20.00) and (4345,22.08) with regard to their best reaches of the fraction 
of $\deltacp$(true) at: 0.77, 0.84, 0.67 and 0.42.}
\label{fig:LvsE}
\end{figure}

Here we are interested to optimize $L$ and $E_\mu$ for the single baseline Neutrino Factory to have CPV discovery at $3\,\sigma$ C.L. 
for different fractions of true values of $\deltacp$ assuming that the true value of $\stheta$ is known. 
The different panels of Fig.~\ref{fig:LvsE} depict the same for four different true values of $\stheta$. 
As expected, this figure clearly shows that the optimization strongly depends on the true value of $\stheta$ chosen. 
In case of large $\stheta \simeq 10^{-1}$, shorter baselines and lower energies are preferred. Even $E_\mu$ as low as 5 to $8\,\mathrm{GeV}$ 
at the Fermilab-Homestake baseline of about $1300\,\mathrm{km}$ is quite close to the optimal choice, which means that the MIND detector 
approaches the magnetized totally active scintillator detector performance of the low energy Neutrino Factory (LENF)~\cite{Geer:2007kn}.
Note that compared to earlier analyses without feed-down effect due to background migration, now too high $E_\mu$ are in fact disfavored 
in the large $\stheta$ case. For the other extreme, $\stheta \simeq 10^{-4}$, baselines between $4 \, 000$ and $5 \, 000$~km are preferred 
with $E_\mu \simeq 20-25$~GeV, which corresponds more to the choice of $L_1$ for HENF, such as the IDS-NF baseline. 
Including the other two panels, the optimal region within each panel moves from the lower left on 
the plots to the upper right as the true value of $\stheta$ decreases. This means that, depending on the choice of true $\stheta$, the optimization results in 
the LENF, the HENF, or an intermediate scenario, and that the low and high energy Neutrino Factories are just two versions of the same experiment 
optimized for different parts of the parameter space. Note that, all the figures presented in this talk have been taken from~\cite{Agarwalla:2010hk} and 
all the simulations have been performed using the GLoBES software~\cite{Huber:2004ka,Huber:2007ji}.

\section{Summary and conclusions}
\label{sec:conclusion}

\begin{figure}[tp]
 \centering
 \includegraphics[width=1.0\textwidth]{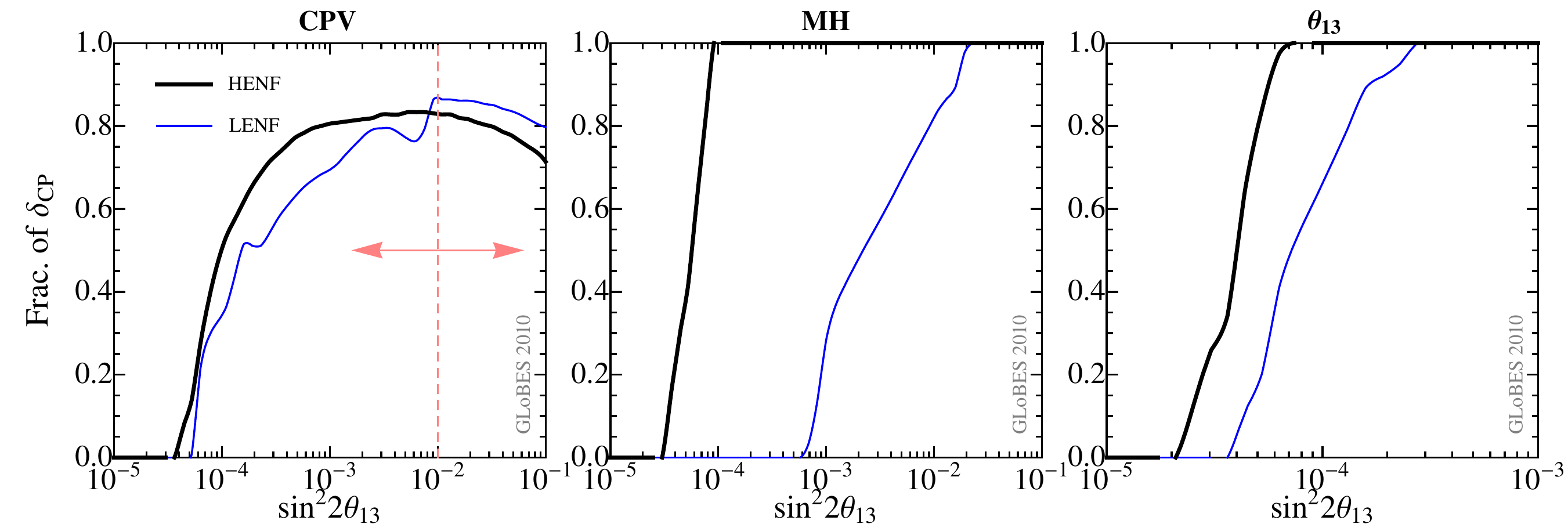}
 \mycaption{CP fractions for which a discovery at $3\sigma$ confidence level is possible as a function of true $\stheta$. From left to right for
 CPV, MH and $\tet$. Here we compare the performances of a single baseline LENF with $E_\mu=10$ GeV, $L=2000\,\mathrm{km}$, and
 a 50 kt MIND detector, and a two baseline HENF with $E_\mu=25$~GeV and a 50 kt MIND detector at each baseline, and
 $L_1=4000\,\mathrm{km}$, and $L_2=7500 \,\mathrm{km}$.}  
\label{fig:LENF}
\end{figure}

In this talk, we have discussed the optimization of the Neutrino Factory based on the most up-to-date analysis of the MIND detector
using response matrices for all the signal and background channels. The lower threshold and higher efficiencies compared to earlier simulations indicate 
that the MIND detector characteristics are getting more similar to the characteristics of the detectors proposed for the LENF 
(\eg, a magnetized TASD). For large $\stheta$, a single baseline Neutrino Factory with $E_\mu$ as low as 5 to 8~GeV and a
 baseline as short as Fermilab to Homestake (about $1300\,\mathrm{km}$) might be sufficient. 
In the case of small $\stheta < 10^{-2}$, however, a two baseline Neutrino Factory with one baseline between about 2500~km and 5000~km, 
the other one at about the magic baseline 7500~km is ideal. The additional information coming from the second baseline which is at the magic
distance (insensitive to $\deltacp$) helps to remove the degeneracies and it enhances the fraction of $\deltacp$ for which CPV discovery is possible. 
Fig.~\ref{fig:LENF} nicely depicts this message where we compare the performance of the optimal single baseline LENF with the optimal 
two baseline HENF for the same MIND detector. It is quite evident from this figure that for $\stheta \gtrsim 10^{-2}$ the low energy 
version can perform all of the required measurements, whereas for smaller values the HENF is clearly better. It suggests that this different 
optimization would be sufficient to compensate for the relative deterioration of performance at large $\stheta$ observed in the traditional HENF. 
 
We conclude that the features of the new detector simulation that the backgrounds are typically reconstructed at lower energies and that the threshold 
is lower have significant impact on the optimization of $E_\mu$. The recent simulation results for the MIND have reduced the performance margin 
between TASD and MIND considerably. Therefore, we suggest that the low- and high-energy Neutrino Factory should not be regarded 
as separate options and they merely correspond to two extreme corners of a common parameter space.

\ack
I am grateful to the conveners of working group 1 for the invitation. I would like to thank Patrick Huber, Jian Tang and Walter Winter for their collaboration.
I would also like to thank Nita Sinha and Paul Soler for insightful discussions during NuFact'11.
I acknowledge the support from the European Union under the European Commission Framework Programme 07 Design Study EUROnu, Project 212372 and the 
project Consolider-Ingenio CUP.


\end{document}